\PassOptionsToPackage{pdfpagelabels=false}{hyperref}
\pdfoutput=1
\documentclass[useAMS,usenatbib]{mnras}

\def\spose#1{\hbox to 0pt{#1\hss}}
\def\ltsim{$\mathrel{\spose{\lower 3pt\hbox{$\sim$}}
        \raise 2.0pt\hbox{$<$}}$\thinspace}
\def\gtsim{$\mathrel{\spose{\lower 3pt\hbox{$\sim$}}
        \raise 2.0pt\hbox{$>$}}$\thinspace}

\def \deg {$^\circ$}
\def \eg {e.g.}

\def \ie {i.e.}

\newcommand{\chandra }{{\em Chandra}}

\newcommand{\fermi }{{\em Fermi}}
\newcommand{\gmrt }{{\em Giant Metrewave Radio Telescope}}
\newcommand{\atca }{{\em Australia Telescope Compact Array}}

\usepackage{graphicx}
\usepackage{amssymb}

\title[A $\mathcal{M}\gtrsim3$ shock in `El Gordo' cluster]{A 
$\mathcal{M}\gtrsim3$ shock in `El Gordo' cluster and the origin of the radio 
relic}

\author[Botteon et al.]{A.~Botteon$^{1,2}$\thanks{E-mail: 
botteon@ira.inaf.it}, 
F.~Gastaldello$^{3}$, G.~Brunetti$^{2}$ and R.~Kale$^{4}$ \\
$^{1}$Dipartimento di Fisica e Astronomia, Universit\`{a} di Bologna, via 
C.~Ranzani 1, I-40127 Bologna, Italy \\
$^{2}$INAF - IRA, via P.~Gobetti 101, I-40129 Bologna, Italy\\
$^{3}$INAF - IASF Milano, via E.~Bassini 15, I-20133 Milano, Italy\\
$^{4}$National Centre for Radio Astrophysics, Tata Institute of Fundamental 
Research, Pune 411007, India \\}
\date{\today}

\date{Accepted XXX. Received YYY; in original form ZZZ}

\pubyear{2016}

\begin{document}
\label{firstpage}
\pagerange{\pageref{firstpage}--\pageref{lastpage}}
\maketitle

\begin{abstract}
We present an X-ray and radio study of the famous `El Gordo', a massive and 
distant ($z=0.87$) galaxy cluster. In the deep (340~$\rm{ks}$) \chandra\ 
observation, the cluster appears with an elongated and cometary morphology, a 
sign of its current merging state. The GMRT radio observations at 
610~$\rm{MHz}$ confirm the presence of a radio halo which remarkably overlaps 
the X-ray cluster emission and connects a couple of radio relics. We detect a 
strong shock ($\mathcal{M}\gtrsim3$) in the NW periphery of the cluster, 
co-spatially located with the radio relic. This is the most distant ($z=0.87$) 
and one of the strongest shock detected in a galaxy cluster. This work supports 
the relic--shock connection and allows to investigate the origin of these radio 
sources in a uncommon regime of $\mathcal{M}\gtrsim3$. For this 
particular case we found that shock acceleration from the thermal pool is still 
a viable possibility.
\end{abstract}


\section{Introduction}

Galaxy clusters are the largest virialized structures in the Universe and form 
via aggregation of less massive systems \citep[\eg][]{press74}. During merger 
events, the intra-cluster medium (ICM) is heated by shocks and is believed to 
become turbulent. Part of the energy involved in these processes is converted 
into non-thermal phenomena that exhibit themselves in the radio band as 
halo and relic emissions \citep[\eg][for a review]{brunetti14rev}. Both radio 
sources are diffuse cluster-scale sources with steep 
spectra\footnote{$S_\nu \propto \nu^{-\alpha}$, with $\alpha$ spectral index.} 
($\alpha\gtrsim1$). Radio halos are generally morphologically connected with 
the 
X-ray emission of the hosting cluster, whereas radio relics are elongated, 
polarized and found in cluster peripheries \citep[\eg][for an observational 
overview]{feretti12rev}. In particular, radio relics are believed to form at 
the gigantic shocks that are generated in major mergers, where cosmic ray 
electrons (CRe) are (re)accelerated \citep[see][for reviews]{bruggen12rev, 
brunetti14rev}. This scenario is supported by the arc-shaped morphologies of 
relics, their high level of polarization and by the fact that an increasing 
number of shocks have been detected at the location of radio relics 
\citep[\eg][]{akamatsu13systematic, bourdin13, shimwell15, eckert16, 
botteon16}. The main difficulty in the understanding of the origin of radio 
relics resides in the low Mach number ($\mathcal{M}\lesssim 3-4$) associated 
with merger shocks. The acceleration efficiency at these weak shocks is indeed 
expected to be small and in several cases it is in tension with the 
observational requirements \citep[\eg][]{markevitch05, macario11, kang12, 
pinzke13, kang14, vanweeren16toothbrush, botteon16}. \\
\indent
ACT-CL J0102--4915 is the most massive cluster detected in the far Universe, at 
a redshift of $z=0.87$ \citep{menanteau12}. For its extraordinary mass of 
$M_{500}\sim8.8\times10^{14}$~$\rm{M_\odot}$ \citep{planck14xxix}, it is also 
known with the 
nickname of `El Gordo'. The cluster was firstly discovered by 
its strong Sunyaev-Zel'dovich (SZ) signal \citep{marriage11} and later 
confirmed through optical and X-ray observations. The system is in a complex 
merger state, as revealed by the double peaked galaxy distribution and the 
elongated morphology of its hot ($kT\sim15$ $\rm{keV}$) ICM 
\citep{menanteau10, menanteau12}. In the radio band, a tenuous halo and a 
double relic system at the cluster NW and SE X-ray boundaries were discovered 
\citep{lindner14}. \\
\indent
In this paper we report the discovery of a strong shock associated with a radio 
relic in `El Gordo' cluster. In particular, our joint \chandra\ and \gmrt\ 
(GMRT) analysis provides interesting insights about the origin of the relic. 
Throughout the paper, we assume a concordance $\Lambda$CDM cosmology with $H_0 
= 70$~$\rm{km\,s^{-1}\,Mpc^{-1}}$, $\Omega_{m}=0.3$ and $\Omega_{\Lambda}=0.7$, 
in which $1'' = 7.713$~$\rm{kpc}$ at the cluster redshift ($z=0.87$). Reported 
uncertainties are 68\%, unless stated otherwise.

\section{Observations and data reduction}

\begin{figure*}
 \centering
 \includegraphics[width=.45\textwidth]{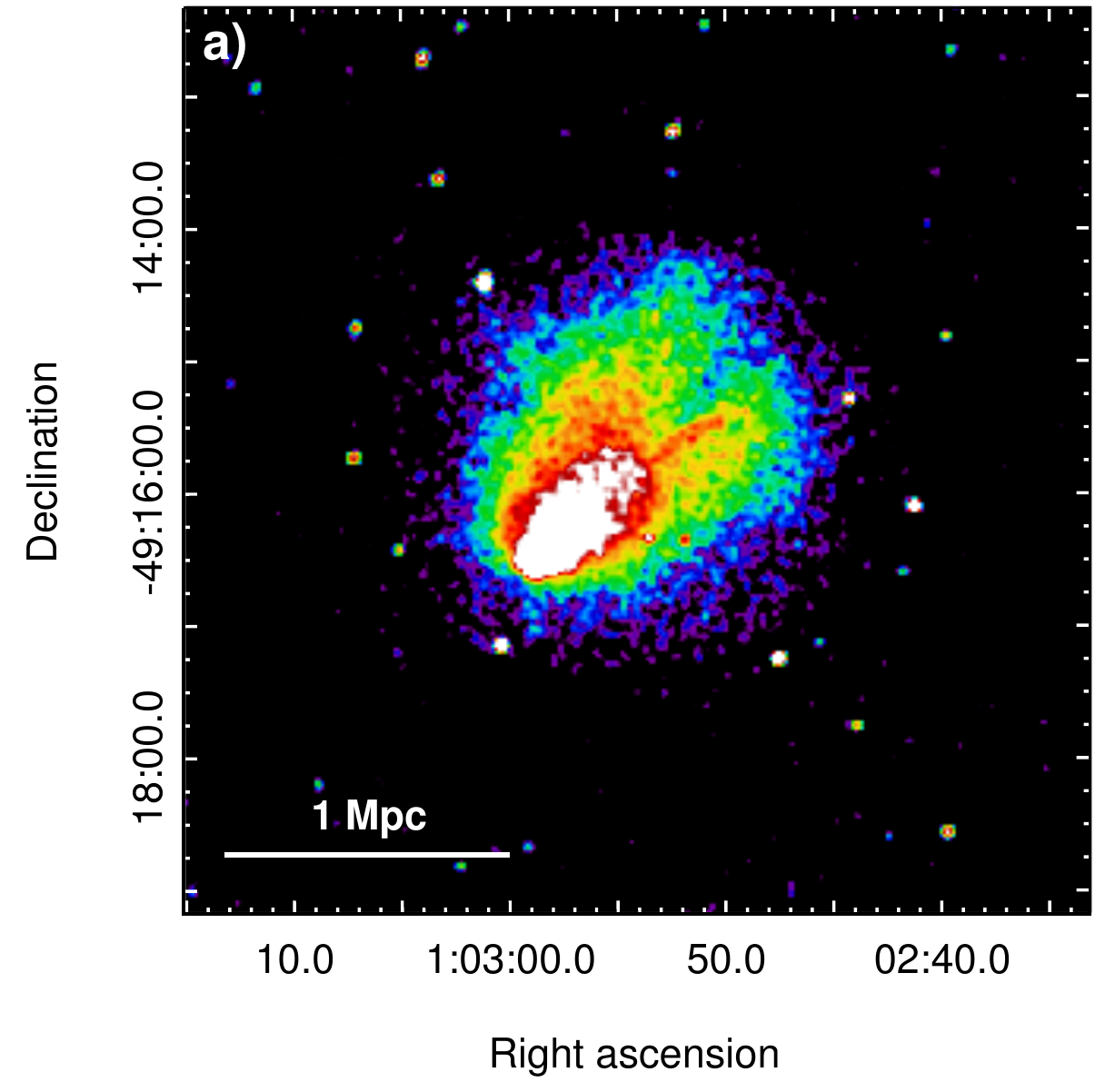}
 \includegraphics[width=.45\textwidth]{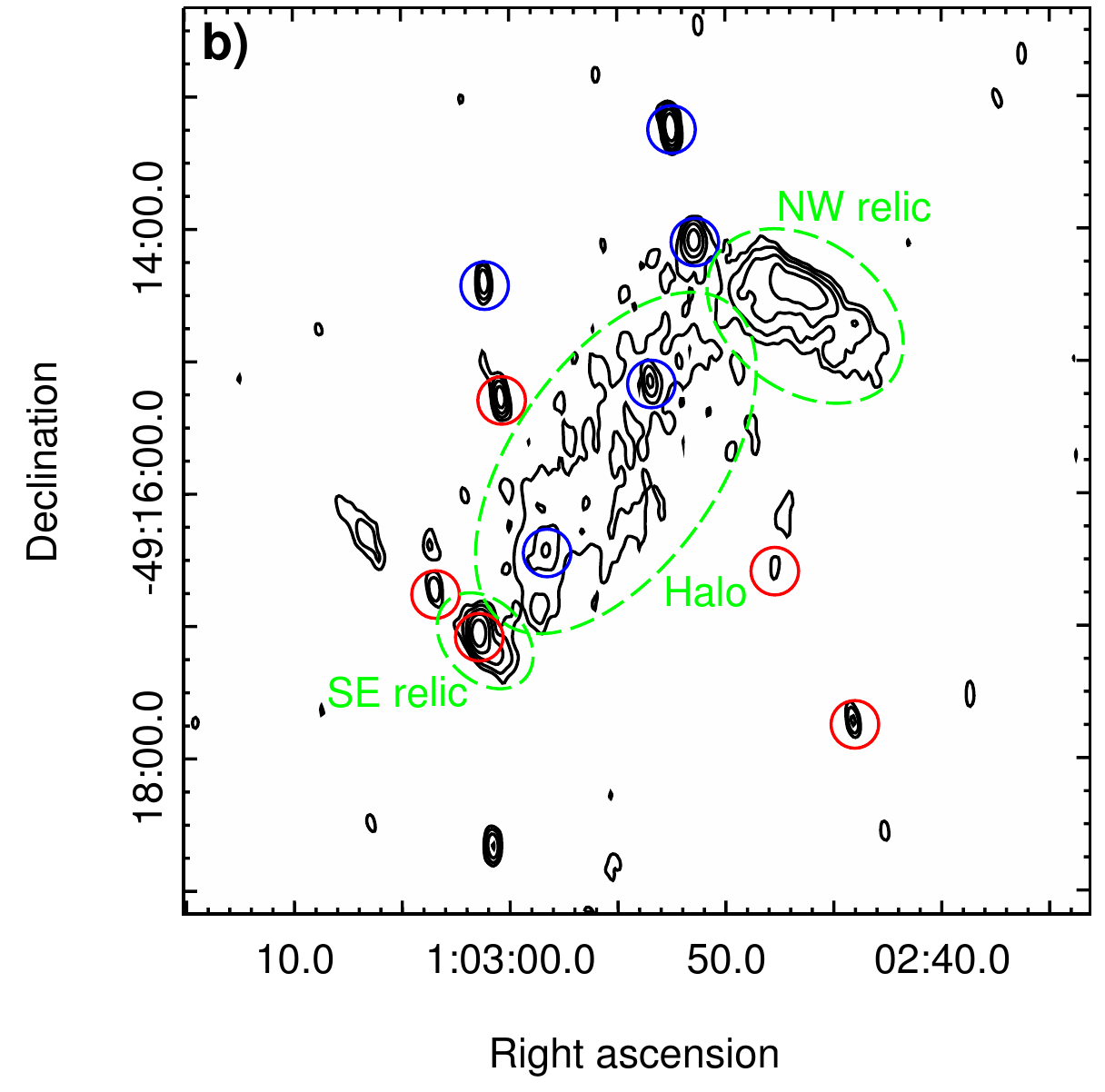}\\
\vspace{1cm}
 \includegraphics[width=.72\textwidth]{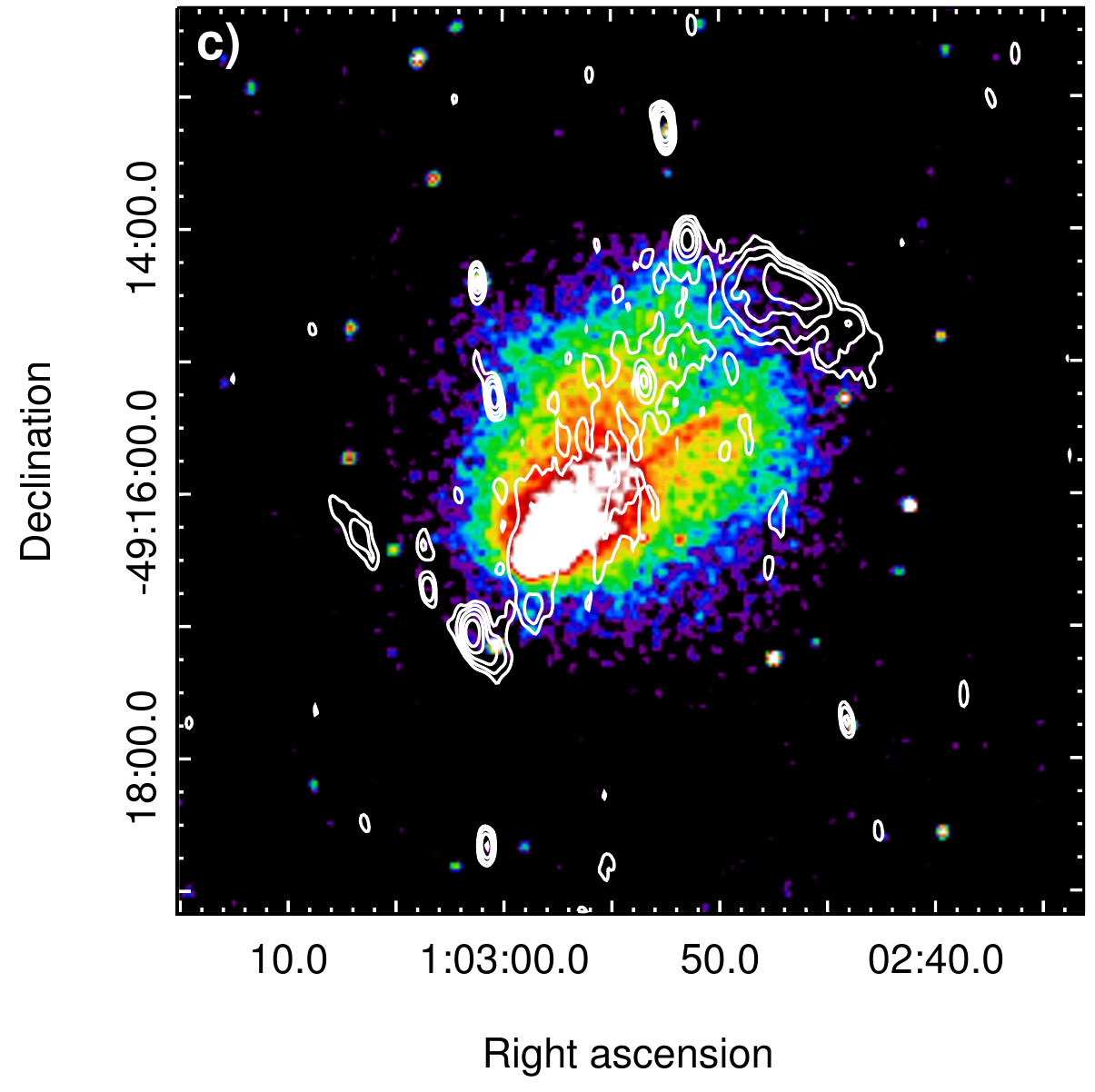}
\caption{`El Gordo' galaxy cluster. \textit{a}) \chandra\ 0.5-2~$\rm{keV}$ 
band exposure-corrected image smoothed on a $3''$ scale. \textit{b}) GMRT 
610~$\rm{MHz}$ radio emission at a resolution of $11'' \times 4.8''$. The 
$1\sigma$ noise level is 50~$\rm{\mu Jy\,b^{-1}}$, contours are drawn at 
levels of $3\sigma \times (-1,1,2,4,8,16)$. Circles denote the compact radio 
sources identified in \citet{lindner14} as cluster members (red) or not 
(blue). \textit{c}) In the \chandra/GMRT comparison the location of the radio
relics at the cluster X-ray boundaries and the spatial connection between the 
halo and the northern X-ray tail are evident.}
 \label{fig:gordo_cluster}
\end{figure*}

\subsection{X-ray data reduction}

`El Gordo' was observed three times (ObsID: 12258, 14022, 14023) with \chandra\ 
in ACIS-I configuration and \texttt{VFAINT} mode for a total exposure time of 
360~$\rm{ks}$. We carried out the standard data reduction by using 
\texttt{CIAO 4.7} and \chandra\ \texttt{CALDB 4.6.9}. In particular, 
soft proton flares were inspected analyzing the light curves extracted from the 
S3 chip in the 0.5-2~$\rm{keV}$ band for each ObsID and removed using the 
\texttt{lc\_clean} routine. We then used the \texttt{merge\_obs} task to make 
the final 0.5-2~$\rm{keV}$ cleaned image (340~$\rm{ks}$) shown in 
Fig.~\ref{fig:gordo_cluster}a. \\
\indent
We created a single exposure-corrected point spread function (PSF) map 
with 
minimum size for the merged image by combining the PSF and exposure maps of 
each 
ObsID. Once the PSF of the instrument is known, the \texttt{wavdetect} task 
allows to identify discrete sources in the surface brightness (SB) image of the 
cluster. These were detected using wavelet radii of 1, 2, 4, and 8 pixels, 
confirmed by eye and excluded in the SB profile analysis. In order to create a 
single background image, the \texttt{reproject\_event} task was used to match 
the background templates to the corresponding event files for every ObsID. This 
single background image was normalized by counts in the band 9.5-12~$\rm{keV}$ 
and subtracted during the SB analysis. \\
\indent
Dealing with spectral analysis of low SB sources as in the case 
of cluster outskirts requires a detailed treatment of the astrophysical and 
instrumental background emission. In this respect, we modeled the sky 
component due to the 
Galactic emission with two thermal plasmas with $kT_1 = 0.14$~$\rm{keV}$  and 
$kT_2 = 0.25$~$\rm{keV}$ , the cosmic X-ray background with an absorbed 
power-law with photon index 
$\Gamma=1.4$ and the ACIS-I particle background by using the analytical 
approach prosed by \citet{bartalucci14}. Spectra were extracted in the same 
region for every ObsID and simultaneously fitted in the 0.5-11~$\rm{keV}$ 
energy 
band with the package \texttt{XSPEC v12.9.0o}. Since the low X-ray count rate, 
we 
kept the metal abundance of the \texttt{APEC} model, which accounts for the ICM 
thermal emission, fixed at the value of 0.3~$\rm{Z_\odot}$ (solar abundance 
table by \citealt{anders89}) and used Cash statistics during the fits.

\subsection{Radio data reduction}

Archival GMRT 610~$\rm{MHz}$ observations of `El Gordo' (project code 
$22\_001$, PI: R.~R.~Lindner) taken on 26 August 2012 were analyzed using the 
Astronomical Image Processing System (\texttt{AIPS}). The GMRT Software Backend 
was used to record the parallel polarization products RR and LL with a 
bandwidth of 33.3~$\rm{MHz}$ divided into 256 channels. The source 3C48 was 
used for flux and bandpass calibration and the calibrator $0024-420$ was used 
for phase calibration towards the target. The total on-target observing time 
was 170 minutes. Standard steps of flagging (excision of bad data) and 
calibration were carried out. The resulting calibrated visibilities towards the 
target were split and used for imaging. A few rounds of phase-only 
self-calibration and a round of amplitude and phase self-calibration were 
carried out to improve the sensitivity of the image. The final image with 
visibilities weighted according to \texttt{ROBUST 0} in the task \texttt{IMAGR} 
and resolution $11'' \times 4.8''$ (position angle $4.8$\deg) is presented in 
Fig.~\ref{fig:gordo_cluster}b. The image was corrected for the GMRT primary 
beam using the task \texttt{PBCOR}. The off-source noise level is 
50~$\rm{\mu Jy\,b^{-1}}$ and a 10\% error on the absolute flux calibration 
scale was assumed. \\
As a preliminary result of new GMRT data, we also used observations 
taken at 327~$\rm{MHz}$ (project code $25\_023$, PI: R.~Kale) to perform 
spectral analysis. The complete analysis of the new radio dataset will be 
presented in the forthcoming paper (Kale et al., in preparation).

\section{Results}

\subsection{X-ray/radio analysis}\label{ch:x-radio}

`El Gordo' X-ray emission remarkably recalls the famous `Bullet' cluster 
\citep{markevitch02bullet}: a dense cool core ($kT\sim 6$~$\rm{keV}$) is
moving in the SE-NW direction producing a prominent cold front 
\citep{menanteau12} which is expected to follow a shock wave
\citep[\eg][]{vikhlinin01a, markevitch02bullet}. The cluster is elongated along 
the merger direction and presents a couple of X-ray tails that give to the 
system a comet-like morphology (Fig.~\ref{fig:gordo_cluster}a). \\
\indent
Our 610~$\rm{MHz}$ radio image of `El Gordo'  recovers extended emission better 
than previously done by \citet[Fig.~2 and 15]{lindner14} as we considered 
baselines down to 0.2~${\rm k\lambda}$ (instead of 0.48~${\rm k\lambda}$). This 
allows to study the morphology of the diffuse sources in more detail. In our 
image shown in Fig.~\ref{fig:gordo_cluster}b, the prominent and elongated radio 
halo connects a pair of radio relics, located in opposite directions at the NW 
and SE edges of the cluster X-ray emission (Fig.~\ref{fig:gordo_cluster}c). The 
strongest part of the halo coincides with the disrupted cluster core, whereas a 
radio tail appears to remarkably follow the northern tail visible in the 
X-rays (Fig.~\ref{fig:gordo_cluster}c). \\
\indent
Our work is focused on the NW radio relic, whose flux densities at 610 
and 327~$\rm{MHz}$ are $F_{610} = 27.5\pm2.8$~$\rm{mJy}$ and $F_{327} = 
64.6\pm6.6$~$\rm{mJy}$, respectively. These result in a radio spectral index 
$\alpha=1.37\pm0.20$. The flux density of the relic at 2.1~$\rm{GHz}$ measured 
by \citet{lindner14} with the \atca\ implies $\alpha\sim1.5$ from 2100 to 
327~$\rm{MHz}$, which is consistent with what we estimated in the narrower 
frequency range. Nevertheless, we will use the spectral index from 610 to 
327~$\rm{MHz}$ since it is taken from two high sensitivity images obtained from 
GMRT observations with matched inner-\textit{uv} coverage (\textit{uv}$_{min}$ 
= 0.2~${\rm k\lambda}$).

\begin{figure}
 \centering
 \includegraphics[width=\hsize]{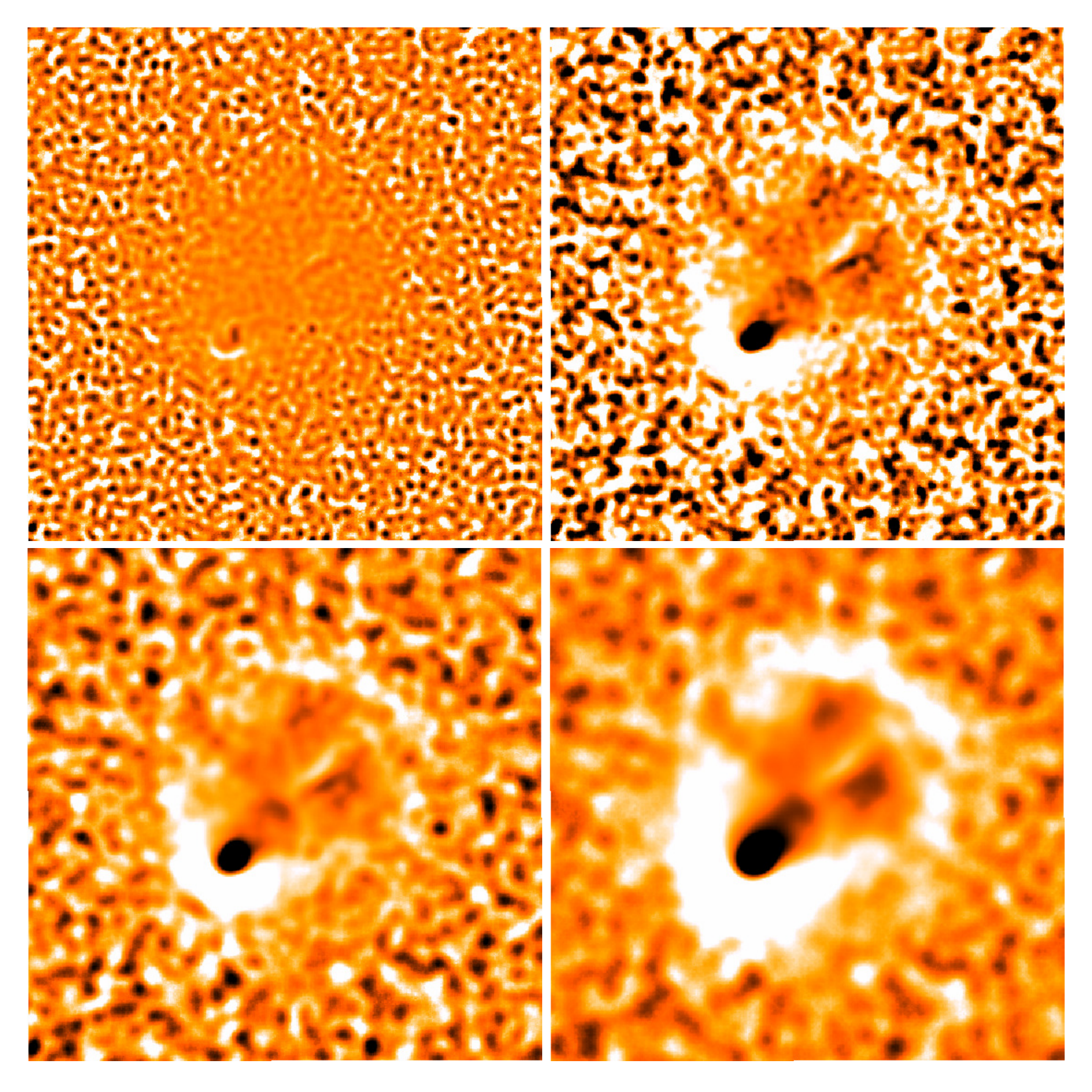}
 \caption{Unsharp-masked \chandra\ images for `El Gordo' cluster created by 
subtracting images convolved with Gaussians with $\sigma_1$ and 
$\sigma_2$ and dividing by the sum of the two. From top left panel in clockwise 
order $(\sigma_1,\sigma_2) = (3'',5''), (3'',20''), (7'',30''), (5'',20'')$.}
 \label{fig:un-mask}
\end{figure}

\begin{figure}
 \centering
 \includegraphics[width=\hsize]{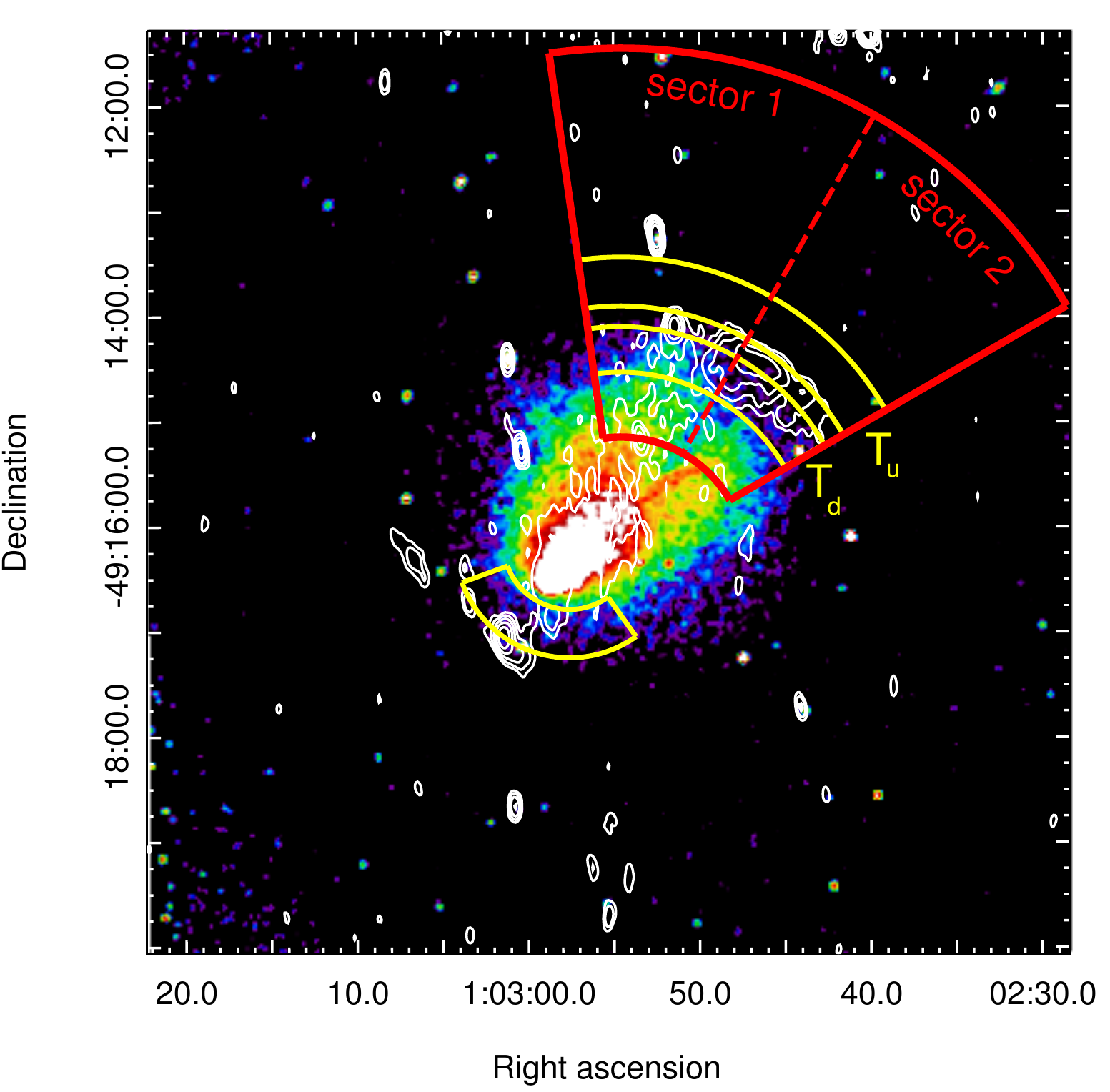}
 \caption{Radio/X-ray overlay of `El Gordo'. Red sectors delineate the surface 
brightness extracting regions. Spectral analysis was performed in the yellow 
sectors. Colors and contours are the same reported in 
Fig.~\ref{fig:gordo_cluster}.}
 \label{fig:sectors}
\end{figure}

\subsection{Relics and shocks}

\begin{figure}
 \centering
 \includegraphics[width=\hsize]{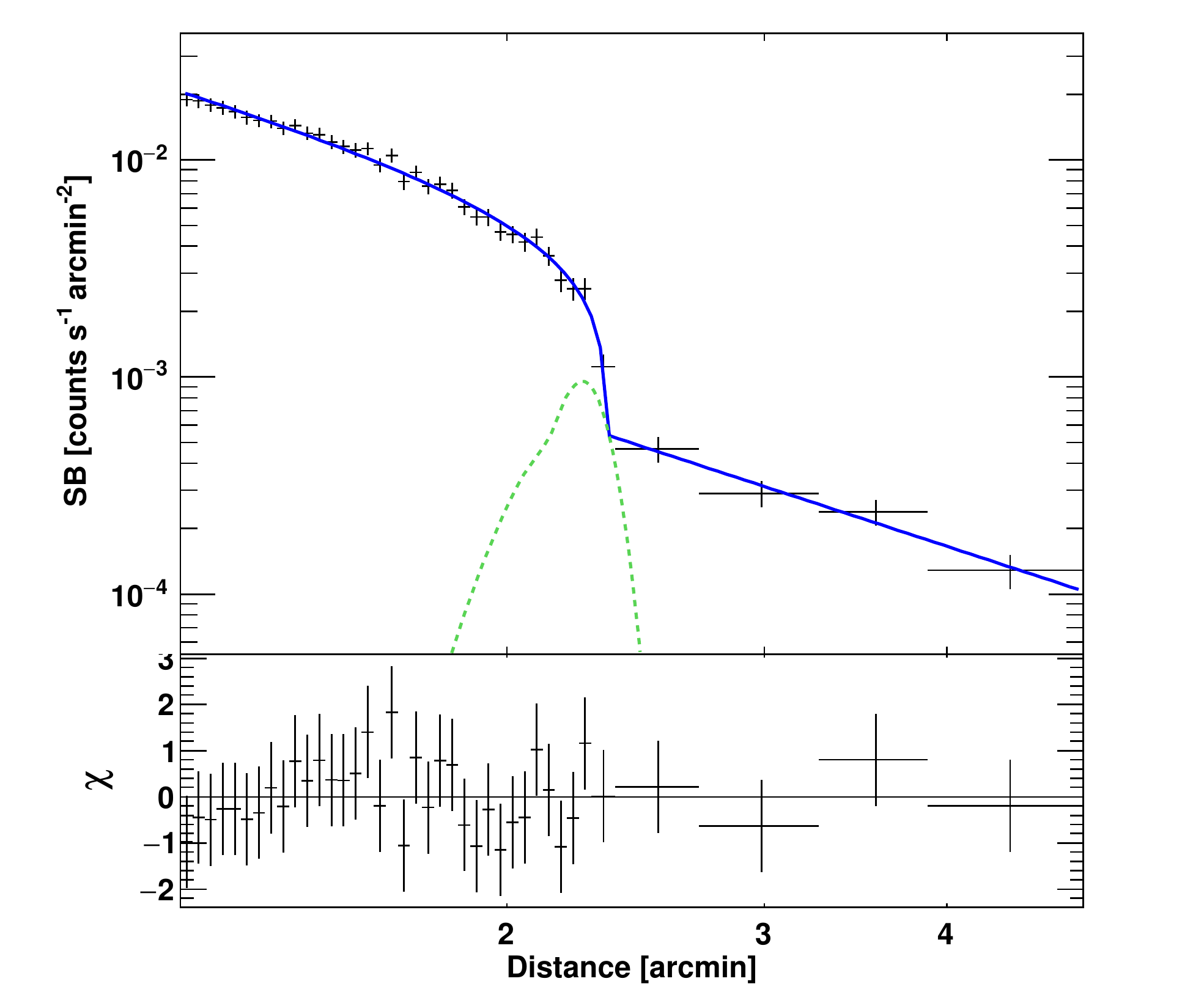}
 \caption{X-ray SB profile in the 0.5-2~$\rm{keV}$ band extracted in sector 
1+2 (Fig.~\ref{fig:sectors}). The data were rebinned to reach 
a minimum signal-to-noise ratio of 7. The green dashed line shows the NW
radio relic brightness profile (in arbitrary units).}
 \label{fig:sb_profile}
\end{figure}

Double relics have been observed in different systems 
\citep[\eg][]{degasperin14} and are believed to form in mergers between 
two clusters with similar mass where diametrically opposite shocks move 
outwards along the merger axis (re)accelerating particles 
\citep{vanweeren11double}. \citet{menanteau12} pointed out the possible 
presence of a couple of shocks by analyzing a $60\:\rm{ks}$ \chandra\ 
unsharp-masked image of `El Gordo'. For these reasons we created the 
unsharp-masked images shown in Fig.~\ref{fig:un-mask} and searched for sharp 
edges in the X-ray SB image, identifying at least one 
discontinuity in the cluster. We used \texttt{PROFFIT v1.3.1} \citep{eckert11} 
to extract the SB profiles in the red sectors shown in Fig.~\ref{fig:sectors}, 
where the NW relic stands out. An underlying broken power-law density profile 
is 
usually adopted to describe the SB in presence of a discontinuity. In the case 
of spherical symmetry, the downstream density $\rho_d$ is higher by a factor of 
$\mathcal{C} \equiv \rho_d/\rho_u$, with $\rho_u$ upstream density, at the 
shock putative distance $r_{sh}$. In formula

\begin{equation}\label{eq:break-pl}
 \begin{array}{ll}
 \rho_d (r) = \mathcal{C} \rho_0 \left( \frac{r}{r_{sh}} \right)^{a_1}, & 
\mbox{if} \quad r \leq r_{sh} \\
\\
 \rho_u (r) = \rho_0 \left( \frac{r}{r_{sh}} \right)^{a_2}, & \mbox{if} \quad r 
> r_{sh}
 \end{array}
\end{equation}

\noindent
where $\rho_0$ is the density normalization, $a_1$ and $a_2$ are the power-law 
indices and $r$ is the radius from the center of the sector (in 
Fig.~\ref{fig:sectors} RA: $+15^\circ.7275$, DEC: $-49^\circ.2724$, J2000). We 
used this density shape to fit the X-ray SB keeping all parameters of the model 
free to vary. \\
We firstly report results concerning sector 1+2 (opening angle (OA): 
$30^\circ-98^\circ$) because it covers the whole extension of the feature 
shown in Fig.~\ref{fig:un-mask} and it gives the maximum SB drop with the best 
statistics (a discussion on the sector choice is presented in 
Section~\ref{ch:syste}). In Fig.~\ref{fig:sb_profile} we report 
the best broken power-law model fit, which is in excellent agreement with data. 
We detect a large SB drop, corresponding to a density compression factor 
$\mathcal{C}=3.4^{+0.4}_{-0.3}$, co-spatially located with the relic. For a 
shock, the Rankine-Hugoniot (RH) jump conditions for a monatomic gas

\begin{equation}\label{eq:mach-from-dens}
 \mathcal{C} \equiv \frac{\rho_d}{\rho_u} = 
\frac{4\mathcal{M}_{\rm{SB}}^2}{\mathcal{M}_{\rm{SB}}^2 + 3}
\end{equation}

\noindent
would lead to a Mach number $\mathcal{M}_{\rm{SB}}=4.1^{+3.4}_{-0.9}$. A 
$\mathcal{M}>3$ shock is quite a rarity in galaxy clusters and so far only two 
of them have been detected ($\mathcal{M}=3.0\pm0.4$ in the `Bullet' cluster, 
\citealt{markevitch06}; $\mathcal{M}=3.0\pm0.6$ in A665, 
\citealt{dasadia16a665}).

\begin{figure}
 \centering
 \includegraphics[width=.73\hsize,angle=-90]{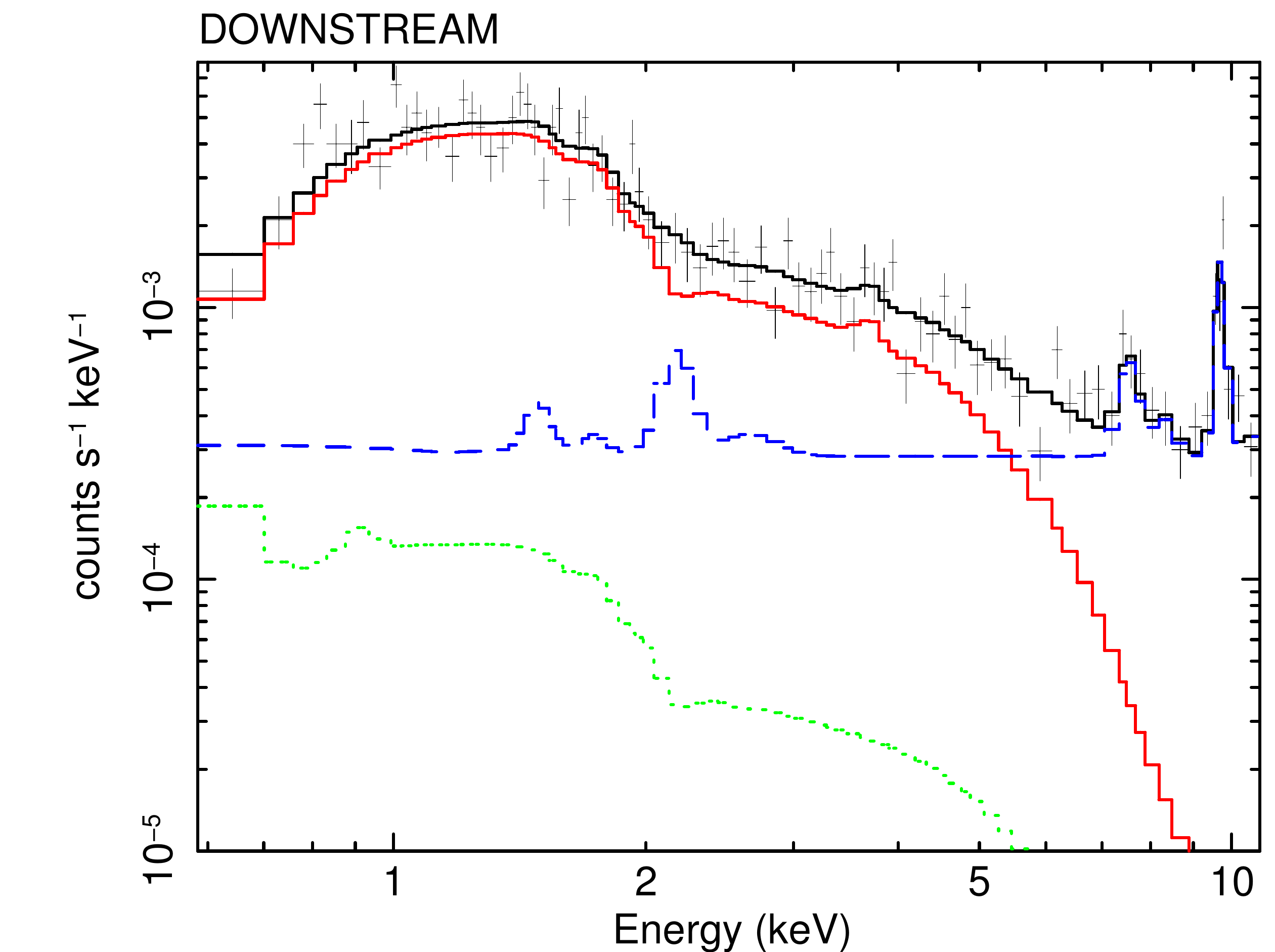} \\
 \includegraphics[width=.73\hsize,angle=-90]{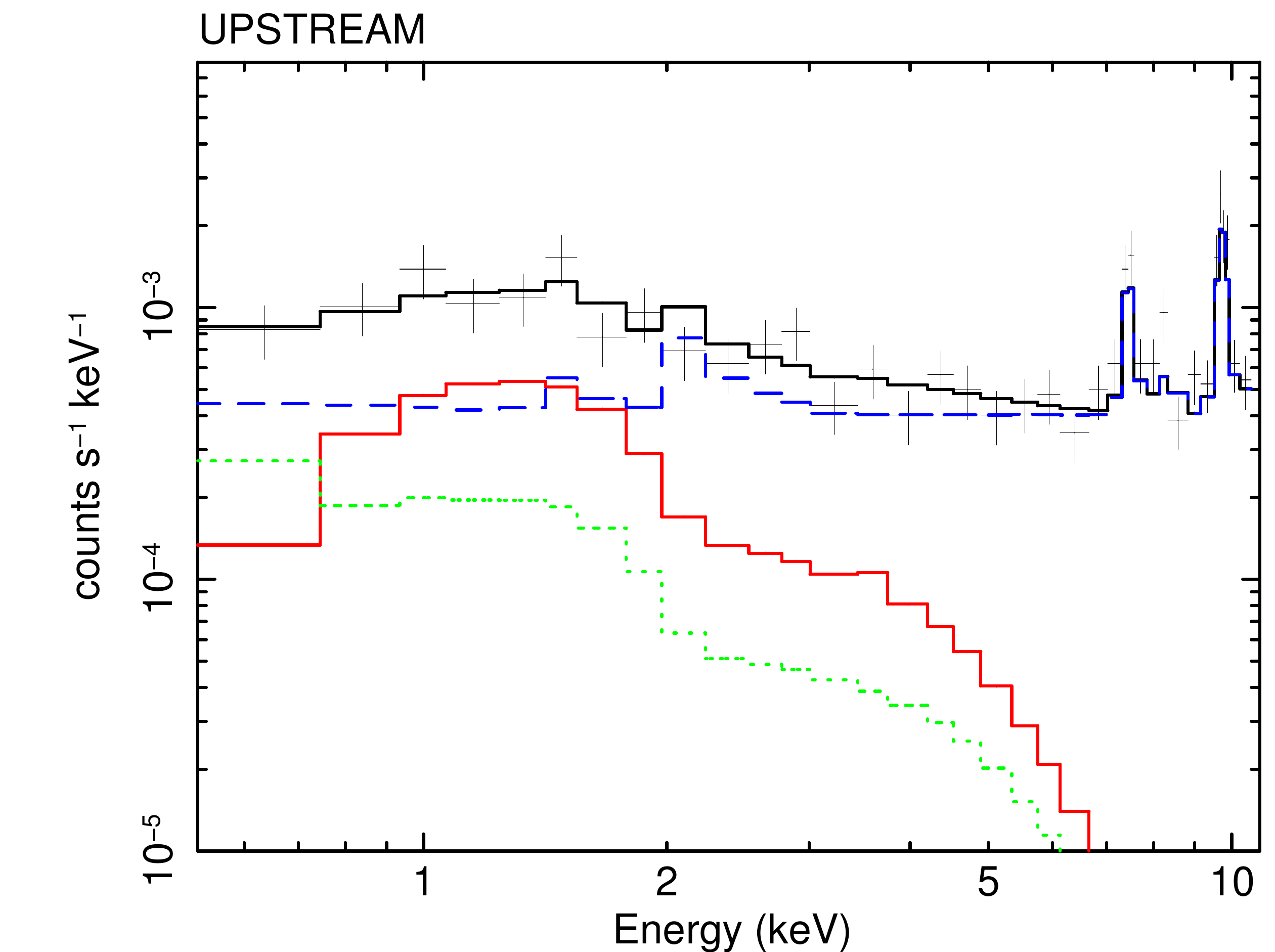}
 \caption{Downstream (top) and upstream (bottom) spectra of sector 1+2. Data 
points are shown in black together with the best fitting model. Different 
colors highlight the model components: the cluster emission (in solid red), the 
particle background (in dashed blue) and the sky background (in dotted green). 
The c-stat/d.o.f. of the fits are 203/168 and 128/115 for the downstream and 
upstream spectrum, respectively. Although spectra were simultaneously fitted, 
only one ObsID was reported in order to avoid confusion in the plot.}
 \label{fig:spectra}
\end{figure}

Shocks heat the downstream gas and this allows to distinguish them from cold 
fronts, other kinds of SB discontinuities found in galaxy clusters with 
inverted temperature jumps \citep[\eg][]{markevitch07rev}. For 
this reason we performed spectral analysis in the yellow sectors shown in 
Fig.~\ref{fig:gordo_cluster}c. Spectra for sector 1+2 are reported in 
Fig.~\ref{fig:spectra}. We found evidence of a very high downstream 
temperature, $kT_d=17.9^{+3.3}_{-2.8}$~$\rm{keV}$, while only a lower limit to 
the upstream one was obtained, $kT_u>6.1$~$\rm{keV}$. In principle, this is not 
enough to confirm the shock nature of the discontinuity but, similarly to the E 
shock in the `Bullet' cluster \citep{shimwell15}, the presence of a cold front 
is very unlikely because it would imply a too high temperature ($kT_u 
> 20$~$\rm{keV}$) at such a large cluster distance. \\
\indent
Although current data do not allow to measure a temperature jump at the 
position of the shock, we can use the lower limit to such a jump to provide 
independent constraints on the shock Mach number. According to RH conditions, 
the upstream and downstream temperature are related by

\begin{equation}\label{eq:mach-from-temp}
 \frac{T_d}{T_u} = \frac{5\mathcal{M}_{\rm kT}^4 + 14\mathcal{M}_{\rm kT}^2 
 -3}{16\mathcal{M}_{\rm kT}^2}
\end{equation}

\noindent
which implies $\mathcal{M}_{\rm{kT}}<2.9$ if we insert the upper $1\sigma$ 
limit of $T_d$ and the lower limit on $T_u$. We anticipate that this value is 
consistent with the Mach number inferred from SB jump once systematic errors 
are taken into account (see the following Section) \\
\indent
A visual inspection of Fig.~\ref{fig:gordo_cluster}a suggests the presence of a 
drop in SB also at the position of the SE relic. A shock in this region is 
expected due to the presence of the radio relic and in analogy with the 
`Bullet' cluster \citep{markevitch02bullet}. However, current data do not allow 
us to characterize statistically the SB drop because of the low X-ray counts in 
this region. Nevertheless we found evidence for high temperature in the 
putative downstream gas, $kT_d=30.1^{+10.5}_{-6.2}$~$\rm{keV}$ in the yellow 
sector in the SE (Fig.~\ref{fig:sectors}), somewhat supporting this 
possibility. Since typical temperatures in cluster outskirts are of a 
few keV, such an high $kT_d$ would likely imply a $\mathcal{M}\gtrsim3.5-4$ 
shock.

\subsubsection{Systematic errors on X-ray analysis}\label{ch:syste}

\begin{table*}
	\centering
	\caption{Results of the SB and spectral fits of the regions shown in 
Fig.~\ref{fig:sectors}. Fits in the $2.0'-4.9'$ radial range were made 
keeping $r_{sh}$ frozen at the best fit value achieved in the wider range.}
	\label{tab:sectors-sb}
	\begin{tabular}{lcccccccc} 
		\hline
		Sector & Radial range & $r_{sh}\:(')$ & $\mathcal{C}$ & 
$\mathcal{M}_{\rm{SB}}$ & $\chi^2 /\rm{d.o.f.}$ & $kT_d$~$\rm{(keV)}$ & 
$kT_u$~$\rm{(keV)}$ & $\mathcal{M}_{\rm{kT}}$ \\
		\hline
		1 & \begin{tabular}{@{}c@{}}$1.2'-4.9'$ \\ 
                              $2.0'-4.9'$\end{tabular} & 
                              $2.359^{+0.006}_{-0.004}$  &
                    \begin{tabular}{@{}c@{}}$3.5^{+0.7}_{-0.5}$ \\ 
                     $3.1^{+0.7}_{-0.5}$ \end{tabular} &
                    \begin{tabular}{@{}c@{}}$>3.0$ \\ 
                     $3.2^{+4.3}_{-0.8}$ \end{tabular} &                     
                    \begin{tabular}{@{}c@{}}20.4/32 \\ 
                     0.9/6 \end{tabular} &   
$18.3^{+4.1}_{-3.2}$ & $6.8^{+10.8}_{-2.8}$ & $1.9-3.4$ \\
		2 & \begin{tabular}{@{}c@{}}$1.2'-4.9'$ \\ 
                              $2.0'-4.9'$\end{tabular} &
                              $2.321^{+0.065}_{-0.041}$ &
                    \begin{tabular}{@{}c@{}}$3.7^{+1.2}_{-0.7}$ \\ 
                     $4.2^{+1.6}_{-0.9}$ \end{tabular} &
                    \begin{tabular}{@{}c@{}}$>3.0$ \\ 
                     $>3.8$ \end{tabular} &                     
                    \begin{tabular}{@{}c@{}}35.4/24 \\ 
                     7.3/2 \end{tabular} &   
$15.8^{+7.9}_{-3.9}$ & $>6.1$ & $<3.1$ \\
		1+2 & \begin{tabular}{@{}c@{}}$1.2'-4.9'$ \\ 
                              $2.0'-4.9'$\end{tabular} &
                              $2.338^{+0.007}_{-0.005}$ &
                    \begin{tabular}{@{}c@{}}$3.4^{+0.4}_{-0.3}$ \\ 
                     $3.4^{+0.5}_{-0.4}$ \end{tabular} &
                    \begin{tabular}{@{}c@{}}$4.1^{+3.4}_{-0.9}$ \\ 
                     $4.1^{+6.7}_{-1.1}$ \end{tabular} &                     
                    \begin{tabular}{@{}c@{}}18.9/34 \\ 
                     5.0/8 \end{tabular} &   
$17.9^{+3.3}_{-2.8}$ & $>6.1$ & $<2.9$ \\
		\hline
	\end{tabular}
\end{table*}

\begin{figure}
 \centering
 \includegraphics[width=\hsize]{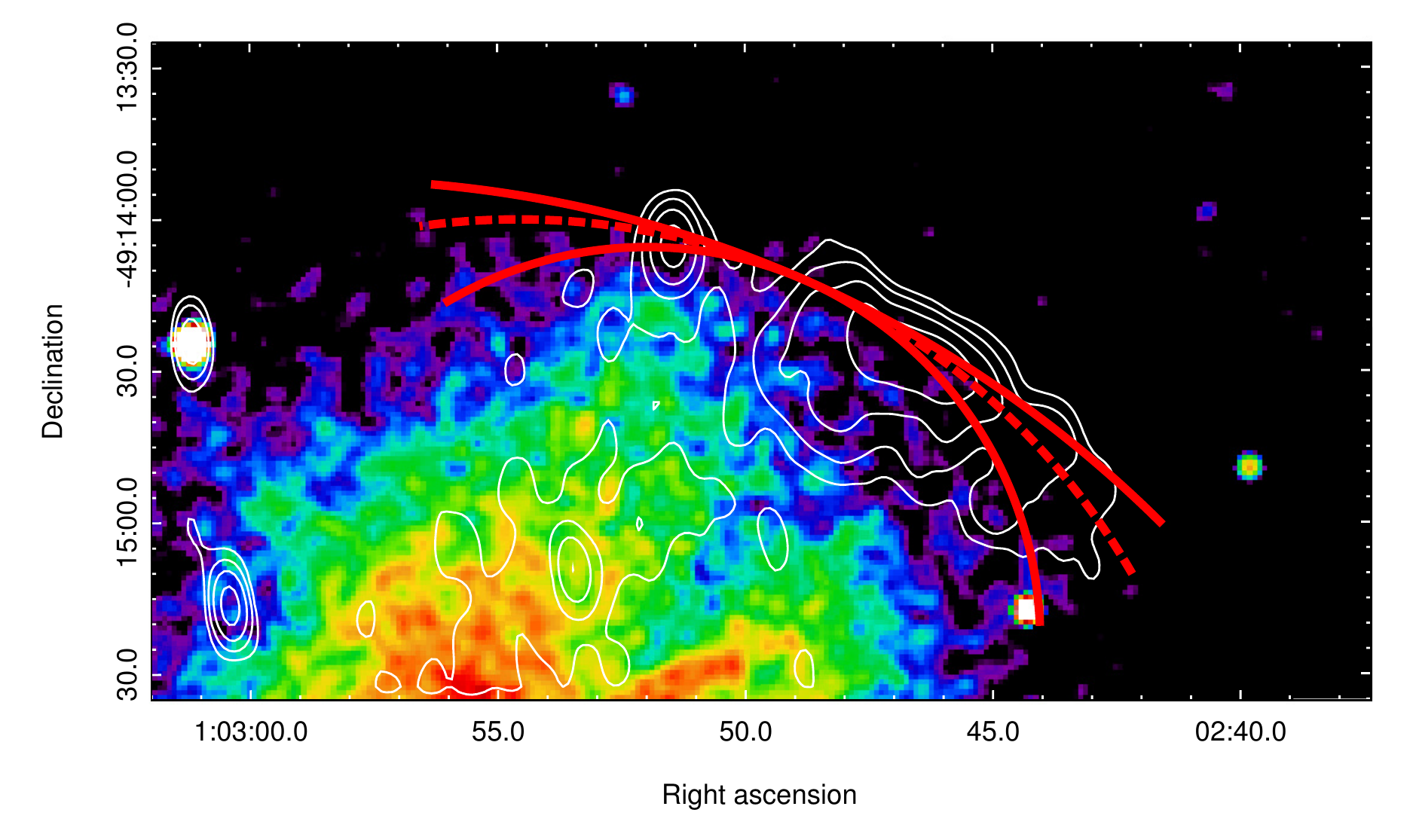} \\
 \includegraphics[width=\hsize,trim={0.1cm 11cm 0.5cm 
0},clip]{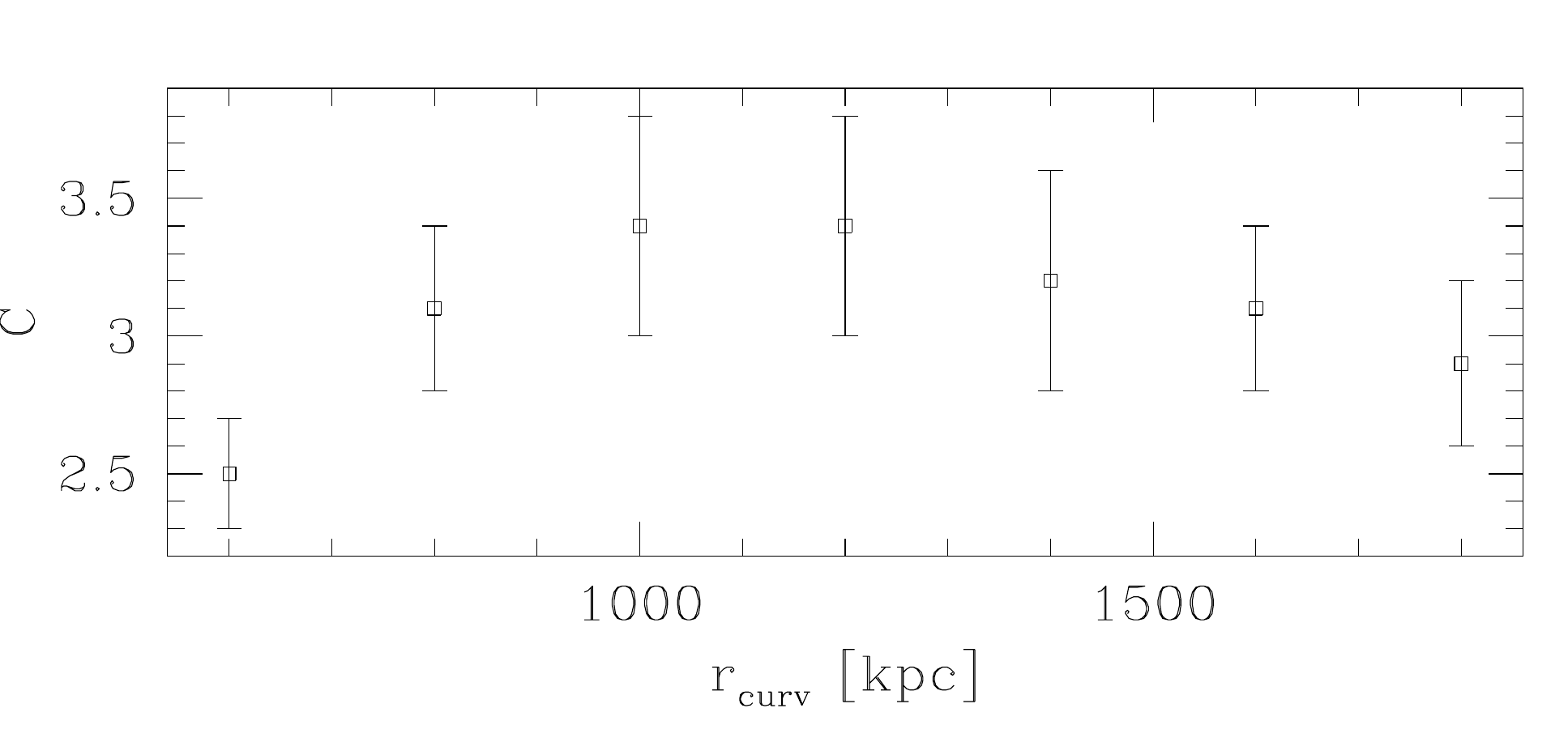}
 \caption{In the top panel we show the difference between the best fit 
curvature radius of $\sim 1$~$\rm{Mpc}$ (dashed line) and the two extreme cases 
with $r_{curv}=600$ and 1800~$\rm{kpc}$ (lower and upper solid lines, 
respectively). In the bottom panel we compare compression factors achieved for 
different values of $r_{curv}$ (see Tab.~\ref{tab:curvature}).}
 \label{fig:r_curv}
\end{figure}

Results in the previous Section are based on measurements obtained for a 
particular sector (1+2). This entirely covers the feature found in the 
unshap-mask images (Fig.~\ref{fig:un-mask}) and allows the best 
characterization of the SB jump due to the statistics of the fit.\\
\indent
We checked the impact due to the choice of the SB extracting region in the 
determination of the NW X-ray discontinuity and the resulting Mach number.
Firstly, we re-performed SB and spectral analysis by splitting the red and 
yellow sectors of Fig.~\ref{fig:sectors} in two sub-regions; the dashed 
line distinguishes between sector 1 (OA: $60^\circ.5-98^\circ$), which 
is oriented in the N direction, to sector 2 (OA: 
$30^\circ-60^\circ.5$), which is in the NW direction and better overlaps the 
relic. In both 
regions, the SB profile is well described by a compression factor $\mathcal{C} 
\gtrsim 3$, implying $\mathcal{M}_{\rm{SB}} \gtrsim 3$. \\
We then repeated the SB analysis by excluding data at $r<2'$ and keeping the 
discontinuity distance frozen at the values found in the $1.2'-4.9'$ radial 
range. Although with a large error, spectral analysis allowed to constrain the 
upstream temperature in sector 1, implying a 68\% confidence interval estimate 
for the Mach number $\mathcal{M}_{\rm{kT}}=1.9-3.4$ (taking into account the 
asymmetric errors on the two temperatures), whereas only lower limits to $kT_u$ 
can be obtained in sectors 1+2 and 2. The results of the fits obtained for the 
three regions are summarized in Tab.~\ref{tab:sectors-sb}. \\
Finally, we checked the variation on $\mathcal{M}_{\rm{SB}}$ in sector 
1+2 due to different shock curvature radii from the best fit value found in 
Fig.~\ref{fig:sb_profile} ($r_{curv} \sim 1$~$\rm{Mpc}$). Results are reported 
in Tab.~\ref{tab:curvature} and the impact of $r_{curv}$ on the shock 
compression factor is presented in Fig.~\ref{fig:r_curv}.\\
\indent
Spectral analysis requires a careful determination of the background sources 
and its systematic uncertainties. In this respect, we varied background 
normalization levels within $\pm1\sigma$ and re-performed spectral fits. We 
achieved results consistent with the reported cluster parameters within 
$1\sigma$. Nonetheless, we highlight that the measurement of high temperatures 
is critical with \chandra\ given its low effective area at energies higher than 
$5$~$\rm{keV}$, in particular the estimated confidence range may not reflect 
entirely the true statistical and systematic error range. \\
\indent
As a final test, a more complex model of a two-temperature 
thermal plasma was adopted to fit the downstream spectra. In this case, the 
high-$T$ component is not constrained while the low-$T$ component gives 
unreasonably low temperatures (\eg\ $kT_{high}>21.2$~$\rm{keV}$ and 
$kT_{low}=2.2^{+4.0}_{-1.2}$~$\rm{keV}$, for sector 1+2). As pointed out in the 
case of the Coma cluster \citep{gastaldello15}, the low-$T$ component mitigates 
the fit residuals at low energy rather than describing a physical condition. 
The high-$T$ spectral component instead supports the presence of a high 
temperature plasma in the downstream region.

\begin{table}
	\centering
	\caption{Impact on the SB profile fits in sector 1+2 due to different 
shock curvature radii. Note that $r_{curv} \sim 1$~$\rm{Mpc}$ in 
Fig.~\ref{fig:sb_profile}.}
	\label{tab:curvature}
	\begin{tabular}{lccc} 
		\hline
		$r_{curv}$~$\rm{(Mpc)}$ & $\mathcal{C}$ & 
$\mathcal{M}_{\rm{SB}}$ & $\chi^2 /\rm{d.o.f.}$ \\
		\hline
		0.6 & $2.5\pm0.2$ & $2.2^{+0.3}_{-0.2}$ & 96.1/48 \\
		0.8 & $3.1\pm0.3$ & $3.2^{+0.9}_{-0.6}$ & 91.5/52 \\
		1.2 & $3.4^{+0.4}_{-0.3}$ & $4.1^{+3.4}_{-0.9}$ & 41.0/45 \\
		1.4 & $3.2^{+0.4}_{-0.3}$ & $3.5^{+1.7}_{-0.6}$ & 24.7/28 \\
		1.6 & $3.1\pm0.3$ & $3.2^{+0.9}_{-0.6}$ & 37.5/27 \\
		1.8 & $2.9^{+0.3}_{-0.2}$ & $2.8^{+0.6}_{-0.3}$ & 43.1/25 \\
		\hline
	\end{tabular}
\end{table}

\subsection{Constraints on the downstream magnetic field}\label{ch:ic}

Relativistic electrons scattering with the cosmic microwave 
background (CMB) photons is expected to produce inverse Compton (IC) 
emission. From the ratio between radio and X-ray emission it is possible to 
constraints the magnetic field in the source region 
\citep[\eg][]{blumenthal70rev}. For this reason we preformed spectral analysis 
in a region enclosing the NW relic and introduced, in addition to the canonical 
thermal model for the ICM, a power-law in the spectral fit. \\
\indent
We assume that the IC spectrum is a power-law with photon index related to the 
synchrotron spectral index via $\Gamma=\alpha+1$. Initially we set 
$\Gamma=2.37$ (see Section~\ref{ch:x-radio}) and kept it frozen in the fit 
while thermal parameters were free to vary. In this case we obtain 
0.5-2~$\rm{keV}$ upper limit to the non-thermal component 
$F_{[0.5-2\:\rm{keV}]} 
< 6.76 \times 10^{-15}$~$\rm{erg\,s^{-1}\,cm^{-2}}$. \\
\indent
The IC measurement is a very complicated issue and can be influenced by several 
factors. We investigated the impact on IC flux estimation by: using the IC 
power-law slope in the range $2.17-2.57$ (consistently with the values 
reported in Section~\ref{ch:x-radio}), keeping the temperature frozen at 13.5 
and 17~$\rm{keV}$ (which covers a range of $kT$ obtained for 
different sector choices in the relic region), varying background 
normalization levels within $\pm1\sigma$ 
and re-performing the fits in the 0.7-11 and 0.9-11~$\rm{keV}$ energy bands. 
In summary, we found upper limits in the range $(2.95-8.51) \times 
10^{-15}$~$\rm{erg\,s^{-1}\,cm^{-2}}$ for the IC flux, representing 
$\lesssim50$\% 
of the thermal model flux in the same energy band (0.5-2~$\rm{keV}$). However, 
in the case of temperature frozen at 17~$\rm{keV}$, the fits formally result in 
a IC detection both for $\Gamma=2.17$ and 2.57; we do not consider these 
detections solid enough due to the systematics related to the 
presence of multi-temperature components and background characterization. We 
further explore the 
possibility of IC detection in Section~\ref{ch:ic-bump}. \\
\indent
By using our upper limits to IC flux, we conservatively 
obtained\footnote{Calculations were obtained using Eq.~32 in 
\citet{brunetti14rev}.} the following lower limits to the downstream magnetic 
field strength: $B\geq3.1$~$\rm{\mu G}$ for $\Gamma= 2.17$, $B\geq4.9$~$\rm{\mu 
G}$ for $\Gamma= 2.37$ and $B\geq7.6$~$\rm{\mu G}$ for $\Gamma= 2.57$. These 
values are in line with other estimates for radio relics 
\citep[\eg][]{bonafede09double, finoguenov10, vanweeren10, 
vanweeren11zwcl0008}. 

\subsection{Acceleration efficiency}

\begin{figure}
 \centering 
 \includegraphics[width=\hsize,trim={3cm 0cm 3cm 
0},clip]{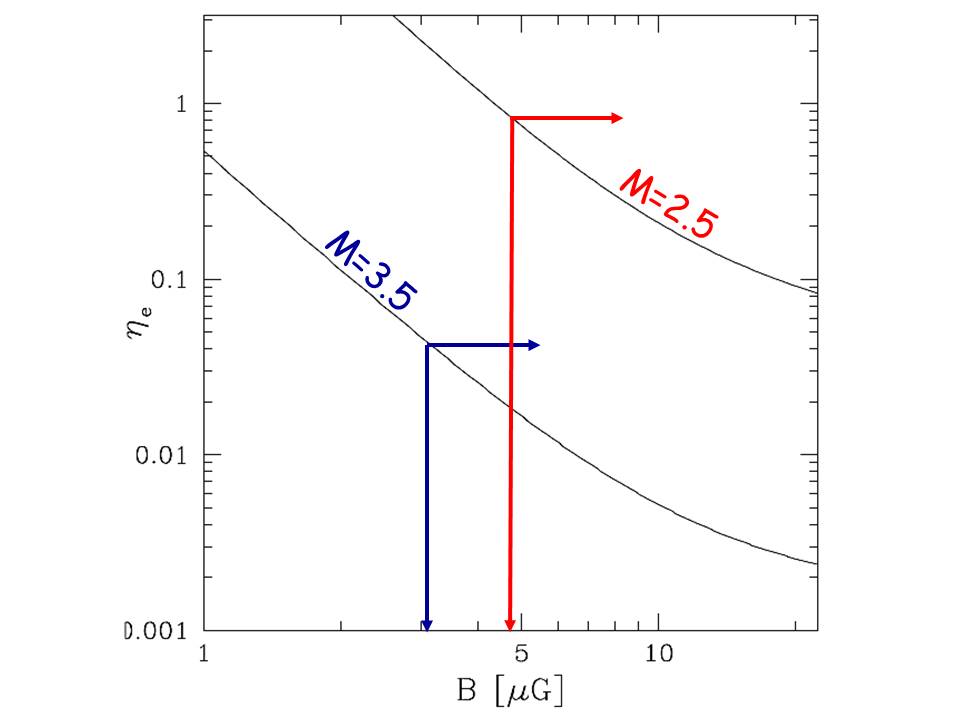}
 \caption{Electron acceleration efficiency versus magnetic field downstream in 
the NW shock in `El Gordo'. Black lines represent efficiencies evaluated for a
Mach number with $\mathcal{M}=2.5$ (top) and 3.5 (bottom). Calculations were 
obtained with $p_{min}=0.1 m_e c$ in Eq.~\ref{eq:n_inj}. Vertical lines denote 
the lower limits on the downstream magnetic field strength achieved from the 
lack of IC emission from the relic.}
 \label{fig:eff}
\end{figure}

The relic--shock connection is nowadays supported by many observational 
studies. Nevertheless, theoretical models of relic formation are challenged by 
the low Mach numbers associated with cluster shocks. In fact, the commonly 
adopted diffusive shock acceleration (DSA) model is severely challenged for 
weak shocks if CRe are accelerated by the thermal pool \citep[\eg][for 
review]{brunetti14rev}. A connected problem is the ratio of cosmic ray protons 
and electrons that would be generated at these shocks and that current 
$\gamma$-ray \fermi\ limits constrain at values that are significantly smaller 
than that in supernova remnants \citep[\eg][]{vazza15efficiency, vazza16}. \\
If the downstream synchrotron luminosity emitted at frequency $\geq \nu_0$ 
originates from electrons in steady state conditions, the bolometric ($\geq 
\nu_0$) synchrotron luminosity that is generated via shock acceleration from a 
shock with speed $V_{sh}$ and surface $S$ can be estimated as

\begin{equation}\label{eq:luminosity}
\int_{\nu_0} L(\nu)\,d\nu \simeq {1 \over 2 } \eta_e \Psi \rho_u 
V_{sh}^3 \left( 1 - {1 \over{\mathcal{C}^2}} \right) {{ B^2 }\over{B_{cmb}^2 + 
B^2}} S
\end{equation}

\noindent
where $\eta_e$ is the efficiency of electron acceleration,   

\begin{equation}\label{eq:psi}
\Psi = {{\int_{p_{min}} N_{inj}(p) E\,dp}\over{\int_{p_0} N_{inj}(p) E\,dp}}
\end{equation}

\noindent
accounts for the ratio of the energy flux  injected in ``all'' electrons and 
those visible in the radio band ($\nu \geq \nu_0$), $p_0$ is the momentum of 
the relativistic electrons emitting the synchrotron frequency $\nu_0$ in a 
magnetic field $B$ and $B_{cmb} = 3.25(1 + z)^2$~$\rm{\mu G}$ accounts for IC 
scattering of CMB photons. The injection spectrum of accelerated CRe is given 
by 

\begin{equation}\label{eq:n_inj}
N_{inj}(p) = (\delta_{inj} +2 ) p^{-\delta_{inj}} \int_{p_{min}}^{p} 
x^{\delta_{inj}-1} N_u(x)\,dx
\end{equation}

\noindent
where $N_u$ is the spectrum of seed particles upstream. In the case of CRe 
acceleration from the thermal pool this is $N_u \propto p^{-\delta_{inj}}$, 
where $\delta_{inj} = 2 (\mathcal{M}^2+1)/(\mathcal{M}^2-1)$, implying a 
synchrotron spectral index $\alpha=\delta_{inj}/2$ in the case of steady state 
conditions \citep[\eg][]{blandford87rev}. \\
\indent
In Fig.~\ref{fig:eff} we report the acceleration efficiency that is necessary 
to explain the radio luminosity observed in the NW relic assuming DSA of 
thermal electrons upstream. \\
We assumed shock Mach numbers $\mathcal{M}=3.5$ and 2.5, in line with the 
values derived from the X-ray analysis. These Mach numbers would imply a 
synchrotron spectrum of the relic $\alpha = 1.18$ and $1.38$, respectively, 
that are in line with radio measurements (Section \ref{ch:x-radio}). 
Calculations were obtained assuming upstream number density and temperature 
$2.4 \times 10^{-4}$~$\rm{cm^{-3}}$ and 6.1~$\rm{keV}$, respectively, and a 
surface of the relic $S=\pi \times 350^2$~$\rm{kpc^2}$. \\
In Fig.~\ref{fig:eff} we also show the lower limits to the magnetic field in 
the relic that are derived from the upper limits to the IC flux assuming the 
two values of the spectral index (Section~\ref{ch:ic}). \\
\indent
Despite we are dealing with a high-velocity shock, $V_{sh} \sim 
4000$~$\rm{km\,s^{-1}}$, we note that the efficiency of CRe acceleration that 
is requested to explain the radio relic is large. This is due to the fact that 
the NW relic in the `El Gordo' is one of the most luminous radio relics known 
so 
far and because, for few $\rm{\mu G}$ magnetic fields, most of the CRe energy 
is radiated via IC emission (due to the high redshift of the cluster). Still, 
contrary to the case of weaker shocks (see \eg\ A754, \citealt{macario11}; 1RXS 
J0603.3+4214, \citealt{vanweeren16toothbrush}; A115, \citealt{botteon16}), we 
conclude that in this case DSA of thermal electrons is still a viable option. 
Indeed for Mach number $\geq 3.5$ the electron acceleration efficiency appears 
energetically viable $\eta_e \leq 0.01$, whereas for $\mathcal{M}\sim 3-3.5$ 
additional mechanisms of pre-acceleration of thermal electrons downstream 
(see \citealt{guo14a,guo14b}) may be required. \\
\indent
The other possibility is that the NW relic is due to shock re-acceleration of 
seeds (relativistic or supra-thermal) electrons. In this case the efficiency 
necessary to explain the radio emission is much smaller simply because the bulk 
of the energy is channelled directly into highly relativistic particles 
(Eq.~\ref{eq:n_inj}, \eg\ \citealt{markevitch05, kang12}). Seeds can be broadly 
distributed in the cluster outskirts where the life time of 100~$\rm{MeV}$ 
electrons is very long (\eg\ \citealt{pinzke13, donnert16arx}) 
or they can be in radio ghost/clouds generated by past AGN activity 
\citep[\eg][]{kang16reacc}. The two possibilities have different predictions 
on the upstream synchrotron emission that in principle can be tested with very 
deep radio observations, but that are well beyond the aim of our paper.

\section{Discussion}

\subsection{Overall considerations}

`El Gordo' is a high redshift ($z=0.87$, \citealt{menanteau12}) and high mass 
($M_{500}\sim8.8\times10^{14}$~$\rm{M_\odot}$, \citealt{planck14xxix}) galaxy 
cluster. It is the most 
distant massive cluster with the brightest X-ray and SZ emission and the 
farthest hosting diffuse radio sources (halo and relics). Our study makes it 
is also the most distant cluster where a shock (with one of the highest 
Mach number) has been detected. \\
\indent
Optical and X-ray observations revealed that `El Gordo' is in a merging state 
\citep{menanteau10,menanteau12}. Recent numerical simulations were able to 
reproduce the overall system properties assuming a nearly head on major merger 
\citep{donnert14, molnar15, zhang15, ng15}. \\
\indent
Double relic systems are expected to trace shocks moving outwards in cluster 
outskirts. So far, studies on `El Gordo' were mainly focused on the SE relic 
located in front of the dense cool core, which is expected to follow a shock in 
analogy with the well know `Bullet' cluster case \citep{markevitch02bullet}. 
However, current X-ray data do not allow to characterize the jump in this 
region 
because of the low statistics. We instead discovered a $\mathcal{M}\gtrsim3$ 
shock spatially coincident with the NW relic. Our detection is based on the 
\chandra\ SB jump. Although with large uncertainties, spectral analysis 
is also consistent with the presence of a strong shock in the ICM. Further 
indications of the shock are given by the unsharp-masked images of 
Fig.~\ref{fig:un-mask}. We also mention the striking similarity between `El 
Gordo' shock/reverse-shock and X-ray morphology with A2146 \citep{russell10, 
russell12} even though the latter is a less massive system 
($M_{500}\sim3.8\times10^{14}$~$\rm{M_\odot}$, \citealt{planck14xxix}) and does 
not host any radio relics at the sensitivity level of current observations 
\citep{russell11}.

\subsection{Hints of inverse Compton emission?}\label{ch:ic-bump}

\begin{figure}
 \centering
 \includegraphics[width=\hsize]{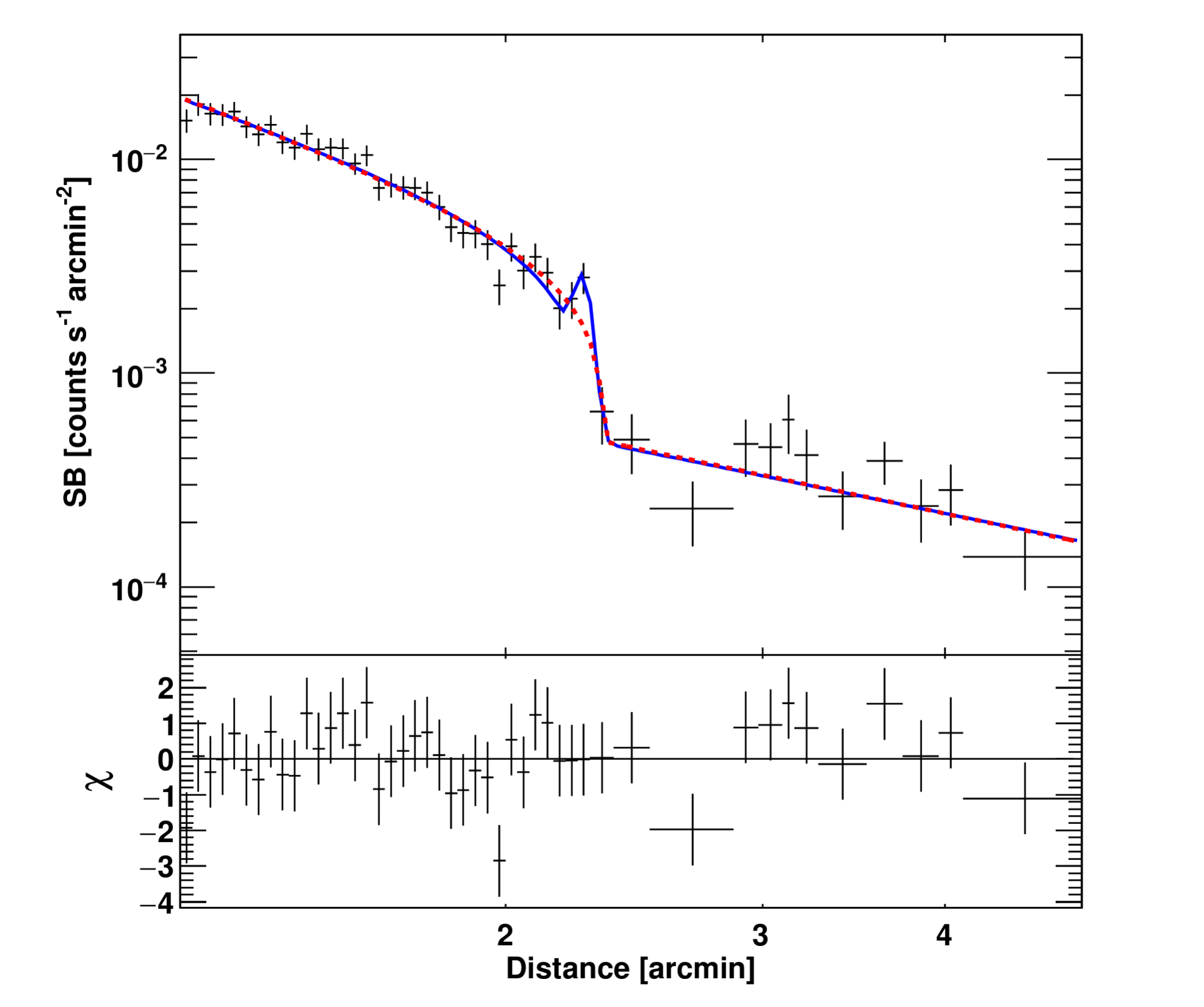}
 \caption{X-ray SB profile in the 0.5-2~$\rm{keV}$ band extracted in a 
region enclosing the NW relic. The broken power-law fit is reported in dashed 
red. The SB excess just before the discontinuity, \ie\ at the relic location, 
for which we speculate a IC origin was modeled by adding a Gaussian component 
to the broken power-law model (solid blue line). Residuals of the 
latter model are displayed in the bottom panel. The data were rebinned to reach 
a minimum signal-to-noise ratio of 3.}
 \label{fig:ic-bump}
\end{figure}

The search for IC emission from galaxy clusters has been undertaken for many 
years with many instruments. However, no confirmed detection has been obtained 
so far. The most famous dispute in this field regards the case of the nearby 
Coma cluster \citep{rephaeli99, rephaeli02, fusco99, fusco04, fusco07, 
rossetti04, wik09, wik11, gastaldello15}. The excellent spatial resolution and 
good spectral capabilities of the \chandra\ satellite allow to minimize the 
contamination from the thermal X-ray emission and open to the possibility to 
search for non-thermal emission also in the soft X-ray band 
\citep[\eg][]{million09}. \\
\indent
`El Gordo' is a perfect target to search for IC emission. It hosts a bright 
radio relic in an external region with low thermal SB and it is at high 
redshift ($z=0.87$), where the equivalent CMB magnetic field strength is large,
$B_{cmb}=3.25(1+z)^2=11.4$~$\rm{\mu G}$. \\
\indent
We used the deep \chandra\ observations to look for IC signatures in the NW 
relic (Section~\ref{ch:ic}). In the case the temperature is frozen at 
17~$\rm{keV}$, the spectral analysis provides a formal detection and a 
significant fraction (up to $40-50$\%) of the X-ray brightness across a relic 
would be contributed by IC emission from the relic itself. This should be 
visible in the X-ray images and profiles across the relic. For this reason we 
extracted a SB profile in the 0.5-2.0~$\rm{keV}$ across a narrow sector (OA: 
$37^\circ.2-67^\circ.5$) containing the NW radio relic. The fit of a 
broken power-law model (Eq.~\ref{eq:break-pl}) in this restricted region 
provides a good description of the SB jump, as shown in Fig.~\ref{fig:ic-bump} 
(dashed red line), leading to $\chi^2 /\rm{d.o.f.} = 48.5/41$. However, a SB 
excess is present in the region of the relic. For this reason we also 
attempt to fit the SB by adding a Gaussian component to the downstream 
power-law. The addition of this Gaussian improves the fit 
(Fig.~\ref{fig:ic-bump}, solid blue line) with $\chi^2 /\rm{d.o.f.} = 40.3/38$.
Slight different sector centers and apertures do not influence this excess. The 
Gaussian component coincident with the relic could represent an excess due to 
IC emission from electrons in the relic. The excess flux associated with the 
Gaussian component is in line with that expected from the spectral analysis. 
The 
combination of this excess with the formal detection of IC emission obtained 
from the spectral analysis (with $kT$ frozen) is tantalizing, however deeper 
observations (\ie\ $>500$~$\rm{ks}$) are required to firmly conclude about this 
possible detection.

\section{Conclusions}

We presented an X-ray/radio study of the famous `El Gordo' cluster located at 
$z=0.87$ focusing on the non-thermal activity in the cluster. \\
\indent
Our GMRT radio observations at 610 and 327~$\rm{MHz}$ confirmed the presence of 
a halo and a system of double relics. These represent the most distant diffuse 
radio sources detected in a galaxy cluster so far. The halo is quite elongated 
in the NW-SE, \ie\ in the merger direction, and remarkably follows the ICM 
emission of the northern X-ray tail. The two relics are found at the boundaries 
of the X-ray emission. We focused on the NW relic which has a synchrotron 
spectral index $\alpha=1.37\pm0.20$ between 610 and 327~$\rm{MHz}$. \\
\indent
The deep \chandra\ observations (340~$\rm{ks}$) allowed us to discover a shock 
at the position of the NW relic. The SB profile 
taken is this region abruptly drops at the relic location. The density 
compression factor $\mathcal{C} \gtrsim 3$ and the high downstream temperature 
provide the indication of a strong shock ($\mathcal{M} \gtrsim 3$) in the ICM. 
This is one of the three strongest shocks detected in galaxy clusters and the 
most distant ($z=0.87$) observed so far. \\
\indent
The detection of a shock co-spatially located with a relic strongly supports 
the relic--shock connection. The NW shock in `El Gordo' cluster allows to study 
particle acceleration in a rare regime of strong shock. We found that DSA of 
thermal electrons is consistent with measured synchrotron spectrum. 
Nonetheless, only shocks with $\mathcal{M} > 3.5$ appear energetically viable 
while for weaker shocks re-acceleration models would be preferred. \\
\indent
The presence of relativistic particles emitting a bright synchrotron relic at 
$z=0.87$ makes `El Gordo' a suitable cluster candidate to search for IC 
emission from the relic. From the X-ray spectral analysis we obtained possible 
hints for IC emission from the relic, however we could not firmly 
conclude the presence of IC excess and conservatively we derived only lower 
limits to the downstream magnetic field that have been used to improve 
constraints on particle acceleration. However, we also found hints of 
an excess in the 0.5-2~$\rm{keV}$ SB profile across the relic region. 
The combination of a possible IC excess in the spectral analysis with the hints 
of excess in the SB is tantalizing and certainly deserves deeper \chandra\ 
observations.

\section*{Acknowledgments}

We thank the anonymous referee for the useful comments on the manuscript. We 
thank Dominique Eckert for his help with the \texttt{PROFFIT} SB 
analysis. The scientific results reported in this article are based on 
observations made by the Chandra X-ray Observatory. We thank the staff of the 
GMRT who have made these observations possible. GMRT is run by the National 
Centre for Radio Astrophysics of the Tata Institute of Fundamental Research. AB 
and GB acknowledge partial support from PRIN-INAF 2014. RK acknowledges 
support through the DST-INSPIRE Faculty Award.

\bibliographystyle{mnras}
\bibliography{library}

\bsp	
\label{lastpage}
\end{document}